\documentclass{aa}  

\usepackage{graphicx}
\usepackage{xcolor}
 \graphicspath{{../theory/}}
\usepackage{txfonts}
\begin{document}

   \title{Implications of the lowest frequency detection of the persistent counterpart of FRB121102}
\titlerunning{Persistent counterpart of FRB121102}

   \author{L Resmi
          \inst{1,2},
  	  J Vink
	  \inst{1},
          \and
          C H Ishwara-Chandra
	  \inst{3}
%         ... \inst{3}\fnmsep\thanks{Just to show the usage
%         of the elements in the author field}
          }
\authorrunning{Resmi, Vink, \& Ishwara-Chandra}
   \institute{Anton Pannekoek Institute, Amsterdam 1098 XH, The Netherlands.
              \\
             \and
             Indian Institute of Space Science \& Techonology, Trivandrum 695047, India.\\
              \email{l.resmi@iist.ac.in}
             \and
             National Center for Radio Astrophysics, Pune 411007, India. \\
%             \email{xyzw@server}
%             \thanks{what to write here}
             }

   \date{Received; accepted}
  \abstract
   {The repeating FRB121102 is so far the only extra-galactic Fast Radio Burst found to be associated with a counterpart, a steady radio source with a nearly flat spectral energy distribution (SED) in centimeter wavelengths.}
   {Previous observations of the persistent source down to $1.6$~GHz has shown no sign of a spectral turn-over. Absorption is expected to eventually cause a turn-over at lower frequencies. Better constraints on the physical parameters of the emitting medium can be derived by detecting the self-absorption frequency.}
   {We used the Giant Metre-Wave Radio Telescope (GMRT) during the period of July to December 2017 to observe the source at low radio frequencies down to $400$~MHz.}
   {The spectral energy distribution of the source remains optically thin even at $400$~MHz, with a  spectral index of $\nu^{-(0.07 \pm 0.03)}$ similar to what is seen in Galactic plerions. Using a generic synchrotron radiation model, we obtain constraints on properties of the non-thermal plasma and the central engine powering it.}
  {We present low frequency detections of the persistent source associated with FRB121102.  Its characteristic flat SED extends down to $400$~MHz. Like Galactic plerions, the energy in the persistent source is carried predominantly by leptons. The emitting plasma has a $B< 0.01$~G, and its age is $> 524 \left(\frac{B}{0.01 {\rm G}} \right)^{-3/2}$. We show that the energetics of the persistent source requires a initial spin period shorter than 36~ms, and the magnetic field of the neutron star must exceed $4.5\times 10^{12}$~G. This implies that the persistent source does not necessarily require energetic input from  a magnetar.}
   \maketitle
%
%-------------------------------------------------------------------

\section{Introduction}
Fast Radio Bursts (FRBs) are extra-galactic %time-domain 
transients of millisecond duration appearing in the radio band of the electromagnetic spectrum (\citet{Lorimer777,Thornton53}, see \citet{Petroff:2019tty} and \citet{Cordes:2019cmq} for recent reviews). Out of the hundreds of FRBs discovered so far, some are found to repeat \citep{2016Natur.531..202S, Amiri:2019bjk, Andersen:2019yex, Kumar:2019htf, Amiri:2019qbv, Fonseca:2020cdd}. Precise localization has been achieved for a handful of FRBs so far \citep{Chatterjee:2017dqg, Ravi:2019alc, 2019Sci...366..231P, 2019Sci...365..565B, Macquart:2020lln}. While the exact mechanism responsible for the coherent radio emission is still unclear \citep{Platts:2018hiy}, FRBs are believed to be associated with neutron stars (NS), particularly magnetars \citep{Popov:2013bia, 10.1093/mnrasl/slu046, Kulkarni_2014,     Katz:2015ltv, Metzger:2017wdz, Nicholl_2017, 10.1093/mnras/stx665, 10.1093/mnras/sty716, Beloborodov:2017juh}. The recent discovery of bright radio bursts from the Galactic magnetar SGR 1935+2154 \citep{Andersen:2020hvz, Bochenek:2020zxn} has confirmed that magnetars can indeed produce coherent radio bursts similar to extra-galactic FRBs, providing a new breakthrough in FRB research.
Except for the X-ray emission associated with this Galactic event \citep{Mereghetti:2020unm}, no FRB has so far shown a transient counterpart in other wavelengths.

The only counterpart associated with an FRB otherwise is the persistent radio source  \citep{Chatterjee:2017dqg,2017ApJ...834L...8M} near the first repeating fast radio burst, FRB121102 \citep{Spitler:2014fla, Spitler:2016dmz}. It is co-located with the FRB position, within a separation less than $12$~mas from the FRB burst position --- corresponding to a projected linear distance of $\sim 40$~pc for $z = 0.19$  \citep{2017ApJ...834L...8M}. Such a close proximity suggests that the two sources may be directly linked. Its radio luminosity is a few orders of magnitude higher than Supernova remnants (SNRs) and pulsar wind nebulae (PWNe) in our Galaxy \citep{2017ApJ...834L...8M}. Therefore, the most popular model for the persistent source is of a nebula powered by a new-born ($< 100$~yr) magnetar \citep{Murase:2016sqo, Metzger:2017wdz, Waxman:2017zme} or a supernova remnant energized by the spin-down luminosity of a young NS \citep{Piro:2016aac}. While no systematic change is seen in the $3$~GHz observations spanning for $150$ days (from April to September 2016), a $10$\% day-scale variability is observed by \citet{Chatterjee:2017dqg}, probably a consequence of scintillation \citep{Waxman:2017zme}.

The persistent source has a flat non-thermal radio spectrum between $1.6$~GHz and $11$~GHz followed by a cut-off \citep{Chatterjee:2017dqg}. Several authors have used the spectral energy distribution (SED) to derive physical parameters of the medium around the FRB assuming a neutron star powered synchrotron nebula for the persistent source \citep{Beloborodov:2017juh, Waxman:2017zme, Murase:2016sqo, Metzger:2017wdz, Dai:2017bjk, Yang:2019xpq}. These calculations use the self-absorption frequency in deriving the physical parameters, which in most cases is assumed at $\sim 1$~GHz,
given that  the lowest frequency observations so far reported for the persistent source was at $1.6$~GHz \citep{Chatterjee:2017dqg}.
To confirm whether the expected synchrotron self-absorption is indeed present, it is important to probe the nature of the SED in lower radio bands. Either detecting a synchrotron self-absorption break frequency, or a lower limit to a break frequency  can be used to derive tighter constraints on the physical parameters of the persistent source. 

We used the Giant Metre-wave Radio Telescope (GMRT) to observe the source in frequencies below $1$~GHz. 
In this paper, we report upgraded GMRT (uGMRT) band-3 ($300 - 500$~MHz) detection of the persistent source at $\sim 200 \mu$Jy level which reveals an optically thin spectrum extending to lower frequencies. While this manuscript was in preparation, another group used a different set of observations in the same bands to present the low frequency SED of the source \citep{Mondal:2020gvn}. The reported fluxes are consistent with our measurements, and further confirm the absence of any systematic change in the source flux. We use a generic synchrotron radiation model to derive constraints on the magnetic field and number density of the non-thermal plasma.  

We present our observations  in section-\ref{sec:obs}, and constraints derived using a synchrotron source model in section-\ref{sec:model}. We summarize our results in section-\ref{sec:end}.
%--------------------------------------------------------------------
\section{GMRT observations}
\label{sec:obs}
Our GMRT observations of the persistent source span epochs from July to December, 2017{\footnote{Under 32\_123,  PI Resmi Lekshmi for legacy-GMRT and under ddtB299, Resmi Lekshmi \& Ishwara-Chandra for uGMRT.}}. In the uGMRT observations, for both band-5 and band-3, we used \texttt{3C147} as both the primary and secondary calibrator. For legacy-GMRT observations we used \texttt{0431+206} as the secondary calibrator for $1390$~MHz observations and \texttt{0410+769} for $610$ and $325$~MHz. Standard calibrators \texttt{3C147, 3C48}, and \texttt{3C286} are used for primary calibration depending on the day of the observation in legacy-GMRT observation.

\begin{figure*}
\begin{center}
\includegraphics[scale=0.35]{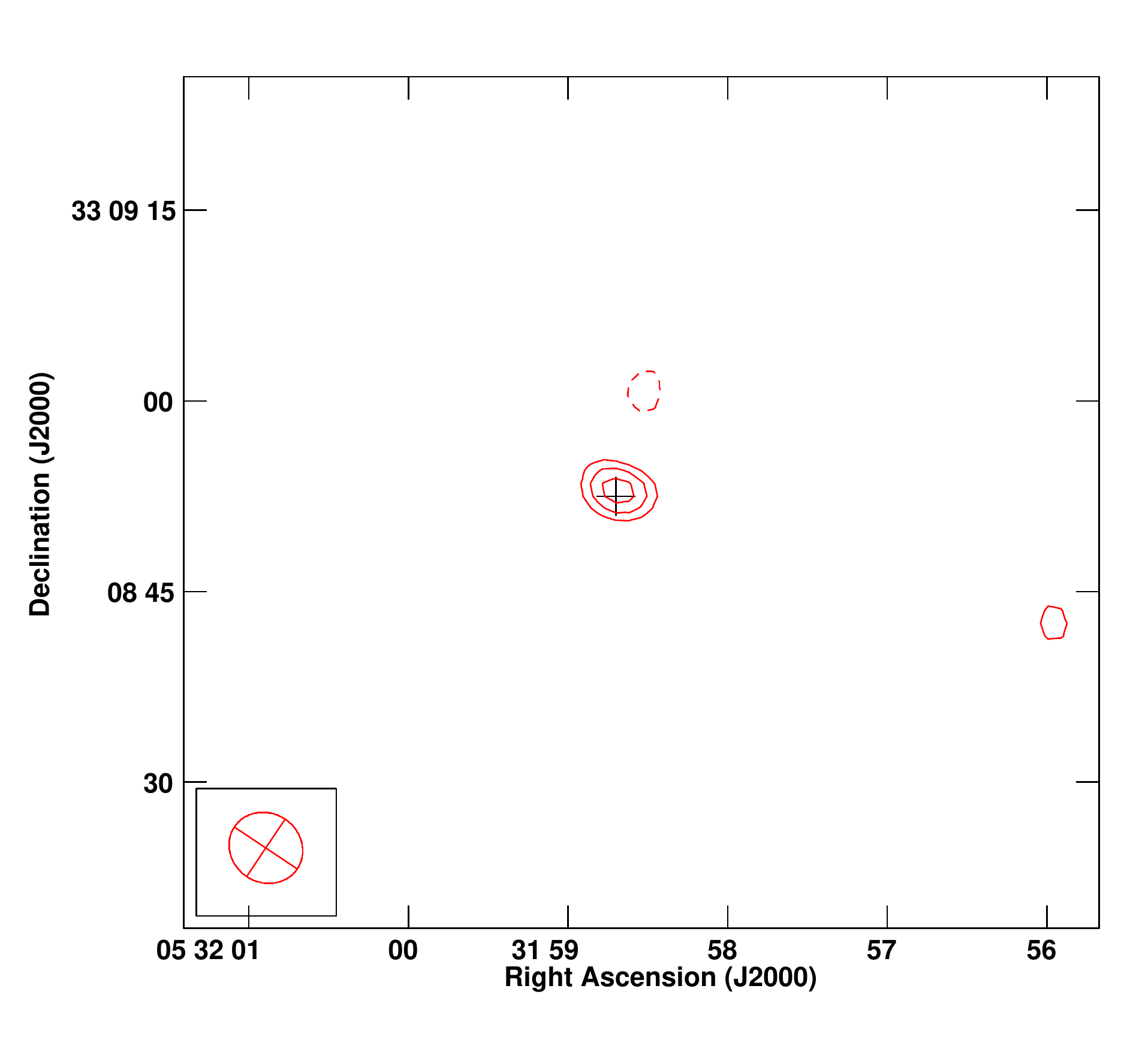}
\includegraphics[scale=0.37]{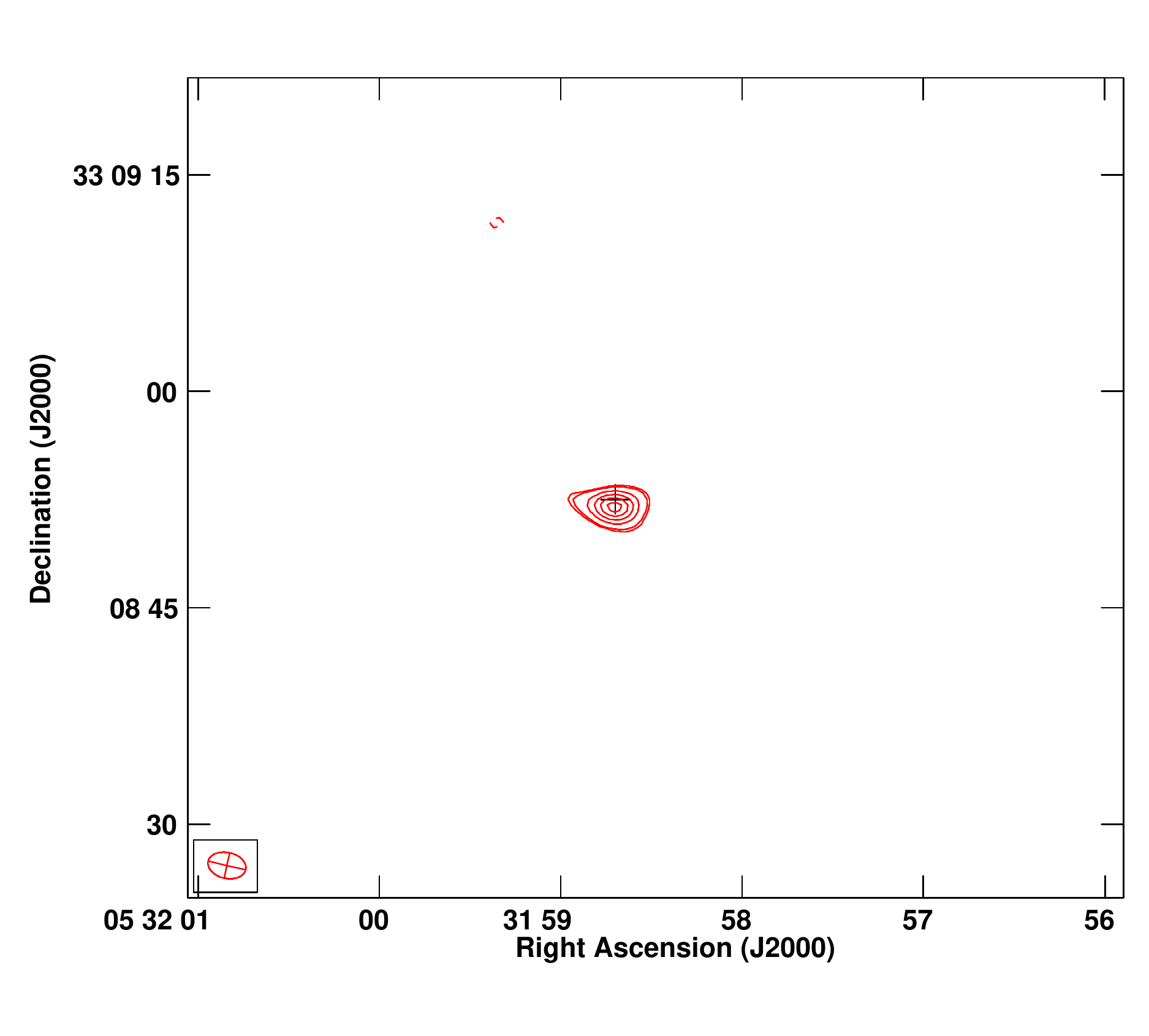}
\caption{uGMRT contour maps of the central portion of the field. The synthesized beam is shown in the lower-left corner. The FRB position from \citet{Chatterjee:2017dqg} is shown as black cross. Left: Band-3 map, with contour levels at $(3,4,5)\times 34 \mu$Jy. Right: Band-5 contours having levels of $(4, 5, 10, 15, 20, 25) \times 13 \mu$Jy}.
\end{center}
\label{fig:maps}
\end{figure*}

Our first observation was in $1390$~MHz on 20 May 2017, followed by in $610$~MHz on 03 Jul 2017 both using the narrow-band $32$MHz correlator. We detected a flux density of  $148.5 \pm 60\mu$Jy in $1390$~MHz and $276.5 \pm 69.0~\mu$Jy in $610$~MHz. We could not detect the source confidently in the $325$~MHz narrow band observation. In order to improve the map quality we further observed the source using the wide-band correlators of the uGMRT in band-3 ($200-300$~MHz) and band-5 ($1050-1450$~MHz) on 16th and 10th Dec 2017 respectively. We detected the source in band-5 with a flux density of $242 \pm 11~\mu$Jy, and in band-3 with $203.5 \pm 33.6 \mu$Jy. The details of all GMRT observations are listed in table-\ref{tab:obsrvn} and  the maps are presented in Fig. \ref{fig:maps}. Our band-5 detections are consistent within the $10$ percent expected variability of the $1.6$~GHz JVLA flux and $1.7$~GHz EVN flux measured by \citet{Chatterjee:2017dqg} and \citet{2017ApJ...834L...8M} respectively.

We used the Astronomical Image Processing System (AIPS) to analyze the narrow band data and a custom-made CASA pipeline to analyze the wide band data.  The fits files were imported to AIPS and the task JMFIT was used to estimate the flux, assuming a two-component intensity distribution, a Gaussian and a flat noise bed, around the position of the FRB. In all maps, the best-fit position of the Gaussian peak is consistent with the reported EVN location by \cite{2017ApJ...834L...8M}, within the GMRT synthesized beam.

\begin{table*}
\caption{GMRT observations of the persistent counterpart of FRB121102}
\label{tab:obsrvn}
\centering
\begin{tabular}{c c c c}
Date of observation & Center frequency & Bandwidth & Flux \\
& (MHz) & (MHz) & $\mu$Jy \\
\hline
$20$ May $2017$ & $1390$& $32$& $148.5 \pm 60$\\
$10$ Dec $2017$ & $1260$& $400$& $241.5 \pm 11.1$\\
$03$ July $2017$ & $610$ & $32$ &$276.5 \pm 69.0$ \\
$16$ Dec $2017$ & $400$& $200$& $203.5 \pm 33.6$\\
\hline
\end{tabular}
\end{table*}
\subsection{Contribution of the associated star-forming region}
Using optical/IR observations of the host galaxy, \cite{Bassa:2017tke} have found that a bright star-forming region encompasses the location of the persistent source. From H-$\alpha$ images, and using the $1.4$~GHz-H-$\alpha$ correlation \citep{2011ApJ...737...67M},  \cite{Bassa:2017tke} estimates $\sim 3\mu$Jy flux at $1.4$~GHz. It is possible that the observed H-$\alpha$ flux is not entirely from the star forming region, which can further reduce its $1.4$~GHz emission for the given correlation. Using a spectral index of $-1$, one of the steepest reported for star-forming regions in low radio frequencies \citep{2016ApJS..227...25R, 2017ApJ...839...35M}, the highest possible flux at $400$~MHz can be calculated as $10 \mu$Jy, well below the detection at band-3. Shallower spectral indices like $-0.5$ will yield a flux of $5 \mu$Jy. Therefore we can safely ignore the contribution of the star-forming region in this analysis. 

   \begin{figure}
   \centering
   \includegraphics[width=\hsize]{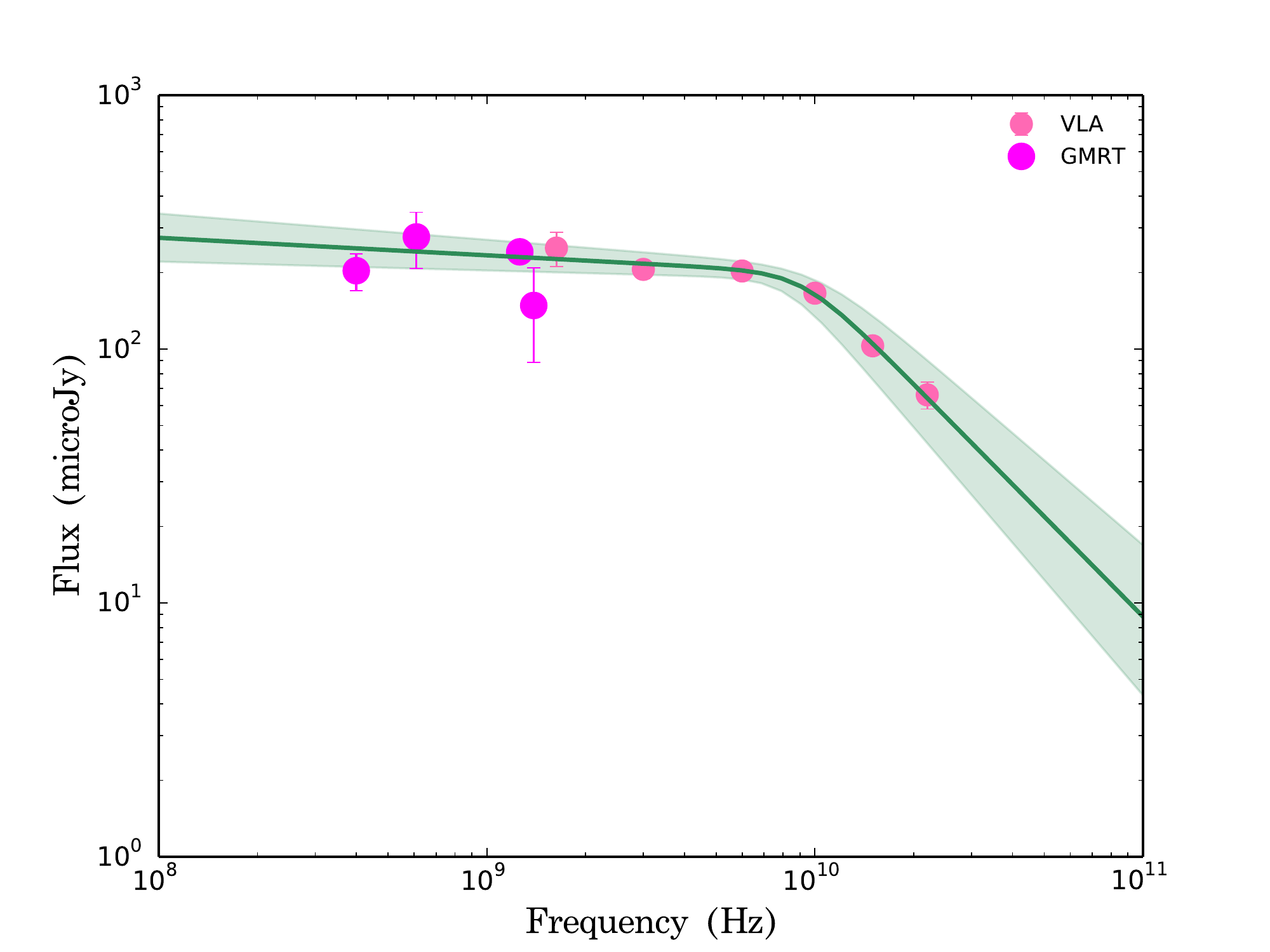}
   \caption{Fit to the near-simultaneous SED of the persistent source by an empirical double power-law model.}
              \label{fig:sed}%
    \end{figure}
\section{Synchrotron spectrum model}
\label{sec:model}
\begin{figure*}
   \centering
   \includegraphics[width=17cm]{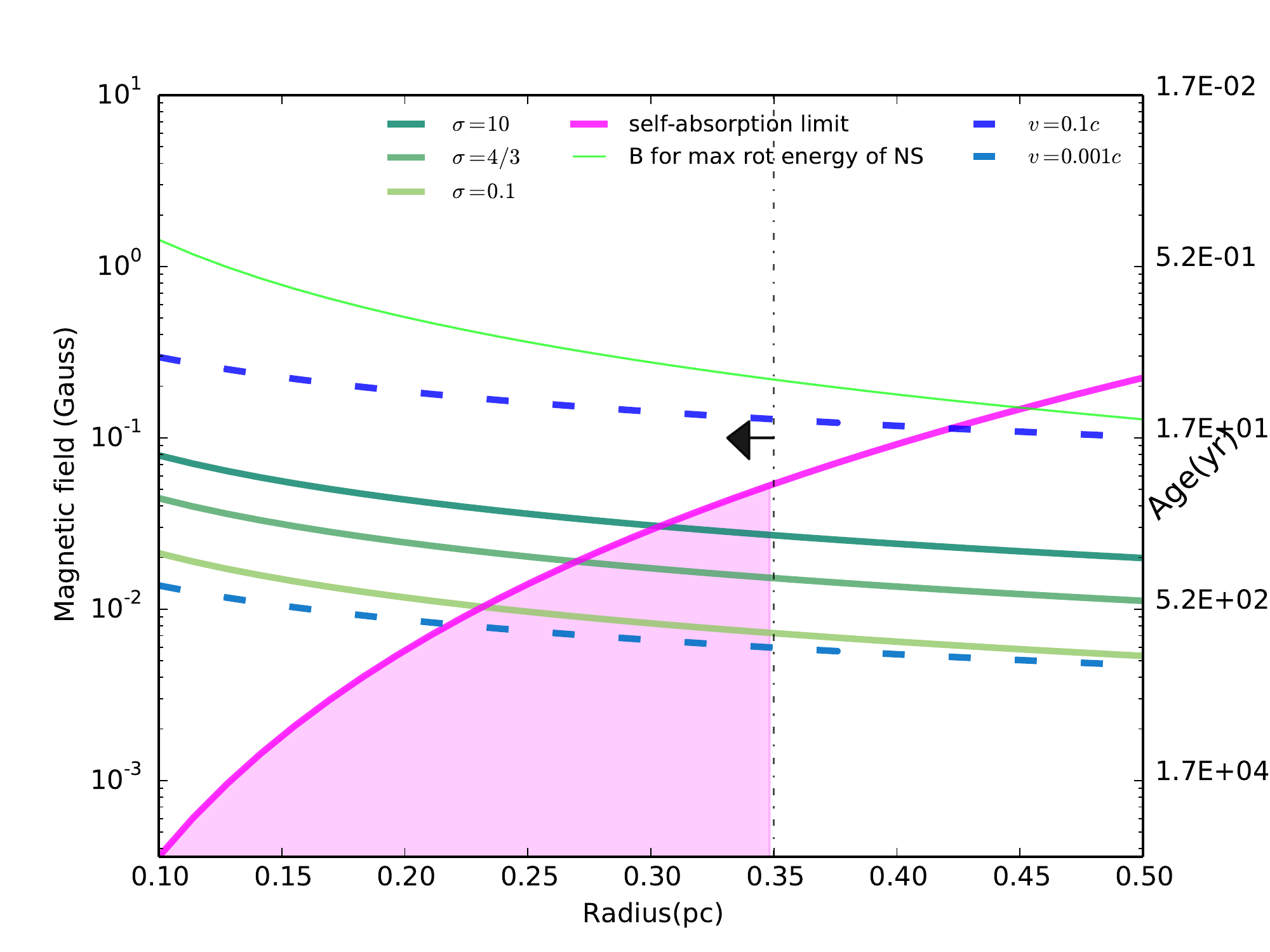}
      \caption{Constraints on the parameters of the emitting plasma from available observations. Minimum energy requirements and optically thin radio SED together restricts the allowed region in the plane of magnetic field ($B$) and radius ($R$). Considering the spectral break at $9.2$~GHz to be due to synchrotron cooling, $B$ can be translated to the age of the source (second y-axis). The magenta solid line and the pink shaded region correspond to self-absorption frequency, $\nu_a \le 400$~MHz. The thick solid lines in shades of green represent $B$ vs $R$ relation resulting from the minimum energy argument for different values of the magnetization parameter $\sigma$ (see text for details). Sensitivity on electron energy spectral index is not high and we have used $p=1.2$ in this calculation. The thin green line corresponds to the highest magnetic field possible, if $10^{52}$~ergs of Magnetar rotational energy is converted to magnetic fields. Dashed lines in shades of blue result from equating the age obtained through synchrotron cooling with $R/v$, where $v$ is assumed expansion velocity. VLBI upper limit on the radius of the source is indicated with a black arrow.}
         \label{fig:constraint}
   \end{figure*}
The uGMRT band-3 detections have confirmed that the spectral energy distribution of the persistent source is optically thin even down to $400$~MHz. Along with the cut-off in the spectrum at higher frequencies observed by the Jansky-VLA, it is possible to arrive at combined constraints for the physical parameters of the emitting plasma.

To proceed, we assume that the emission process is non-thermal synchrotron radiation. The radio SED of the persistent source is flat like that of Galactic plerions, and has a break around $10$~GHz. This break could be related to the acceleration process or due to synchrotron radiative losses. In the latter case, for a continuous injection of electrons, the spectral index increases by $0.5$, much shallower than what is observed for this source. However, instantaneous injection can produce the sharp rollover as we will later show in section-\ref{subsect32}. 
Motivated by this, to begin with we fit the SED with a double power-law spectral model given by, 
\begin{equation}
f_{\nu} = f_0 \left( \left[\frac{\nu}{\nu_b}\right]^{s m_1} + \left[\frac{\nu}{\nu_b}\right]^{s m_2}  \right)^{-1/s},
 \end{equation}
where $f_0$ is the flux normalization, $\nu_b$ is the spectral cut-off, $m_1$ is the asymptotic spectral index for $\nu \ll \nu_b$ and $m_2$ is the same for $\nu \gg \nu_b$. We found that keeping the smoothing index $s < 5$ leads to poor inferences of $m_1$, therefore in the results used below we have assumed $s=5$. The inferred values of other parameters are slightly sensitive to the value of $s$ assumed. Minimizing the $\chi^2$ through a \textit{Levenberg-Marquardt} algorithm, we find that $m_1 = 0.07 \pm 0.03, m_2 = 1.31 \pm 0.2, f_0 = 200.7 \pm 11.3~\mu$Jy, and $\nu_b = 9.2 \pm 1.0$~GHz, leading to $f_{\nu = 9.2{\rm GHz}} = 175~\mu$Jy. The best fitted model is presented in Fig. \ref{fig:sed}.

The difference $\delta m$ in the spectral indices is $\sim 1.25$, very different from what is expected due to synchrotron cooling of continuously injected electrons \citep[$\delta m = 0.5$, e.g.][]{2011hea..book.....L}. Hence, we assume the break to be due to radiative
cooling of an electron/positron population that was
predominantly injected during a time scale that was relatively short compared to the age of the persistent
source. %
Along with the observed absence of self-absorption in low frequencies, in the next section we derive constraints on the emission region assuming it to be powered by a pulsar.
For our interpretation of the spectral properties, it is assumed that 
the flat spectrum is  caused by an intrinsically flat electron/positron injection spectrum.
We note that this is generally the case for pulsar wind nebulae,
but other interpretations may be possible. For example,
for
some extra-galactic radio sources, a flat spectrum could also result from superposition of individual self-absorbed synchrotron components \citep{,1979ApJ...232...34B,1980ApJ...238L.123C}, a possibility we are not considering in this article.

\subsection{Constraints on the physical parameters of the plasma}
\label{subsect31}
In this section, we assume a non-thermal electron distribution, emitting  synchrotron radiation, assumed to result from a uniform magnetic field strength, $B$. High energy electrons are affected by radiative cooling. The optically thin spectrum implies that the photons are not self-absorbed. For a compact source, a limiting value of self-absorption implies a lower magnetic field. On the other hand, energetic considerations can give a relation between $B$ and the radius, $R$, for an observed flux, where $B$ decreases monotonically for increasing $R$. Together, these two arguments, therefore, can provide stringent limits on the $B-R$ plane. In addition, the {\em absence of a} synchrotron cooling break, $\nu_{\rm b}$,{\em above} 
400~MHz
gives constraints on the age of the source for a given magnetic field.

\subsubsection{Self-absorption limit}
First, we describe our derivation of the self-absorption frequency. For a source optically thick to frequency $\nu$, the observed flux is $f_{\nu} = \frac{2 k_B T_B \nu^2}{c^2} \pi \frac{R^2}{d_L^2}$, where $k_B$ is the Boltzmann constant and $c$ is the speed of light. The term $\pi R^2/d_L^2$ represent the solid angle subtended by the emission region of radius $R$ at a distance $d_L$. For a synchrotron source, the brightness temperature $T_B$ can be approximated as $\gamma_a m_e c^2/k_B$, where $\gamma_a$ is the Lorentz factor of the electron whose synchrotron power-spectrum peaks at the self-absorption frequency $\nu_a$. $\gamma_a$ and $\nu_a$ are related through the characteristic synchrotron frequency $\nu_{\rm syn}$. Using, $\nu_a = \nu_{\rm syn}(\gamma_a) = \frac{3 e}{4 \pi m_e c} \gamma_a^2 B$, where $e$ is the elementary charge and $m_e$ is the mass of electron, one can finally obtain the flux at $\nu_a$ to be,
\begin{equation}
f_{\nu_a} = 2.8 \times 10^{-4} \mu Jy \, \left( \frac{R_{\rm pc}}{{d_L}_{Gpc}} \right)^2 \, \left( \frac{\nu_a}{MHz} \right)^{5/2} \, \left( \frac{B}{1 G} \right)^{-1/2}.
\label{selfabseqn}
\end{equation}
Using this equation along with the observed flux at $400$~MHz, we obtain the upper-limit to $B$ as a function of the radius of the plasma. In Fig. \ref{fig:constraint}, this upper-limit is represented by the magenta line. The region in the $B-R$ plane is consistent with the optically thin spectrum down to $400$~MHz is shaded in pink.

\subsubsection{Energetic constraints}
Next, we derive the relation between $B$ and $R$ for for an arbitrary magnetization parameter $\sigma$ defined as the ratio $u_B/u_e$ between the energy in the magnetic field to the non-thermal electrons. This is a generalization of the equi-partition argument. 

The energy density in non-thermal electrons can be written as $u_e \approx \frac{K_e}{2-p} (\gamma_M m_e c^2)^{(2-p)}$, for $p < 2$, where $\gamma_M$ is the maximum electron Lorentz factor, $p$ is the power-law index, and $K_e$ is the normalization of the electron distribution. As the electron distribution is flat for this source, the dominant contribution to $u_e$ comes from electrons at $\gamma_M$ (i.e, in this case the electrons radiating at the break $\nu_b$). This assumption is valid even if the break is due to radiative cooling, as essentially the distribution is bounded within $\gamma_M$. Therefore, we have considered the minimum electron Lorentz factor $\gamma_m \ll \gamma_M$ and ignored it in the equation. In terms of the observed luminosity at the break, the normalization $K_e$ can be written as 
\begin{equation}
K_e = \nu_b L_{\nu_b} \frac{9}{2} \frac{(3-p) (m_e c^2)^{p-1}}{c \sigma_T R^3 B^2 \gamma_M^{3-p}}, 
\label{keeqn}
\end{equation}
where $\sigma_T$ is the Thomson scattering cross-section. We have provided the derivation of this equation in the Appendix-A. 

After substituting for $\gamma_M$ in $u_e$ in terms of $\nu_b$, and since $u_B = B^2/(8 \pi)$, we finally obtain the relation between $B$ and $R$ for a given magnetization parameter $\sigma$ as,
\begin{equation}
B^{7/2} = 8 \pi \sigma {\nu_b}^{1/2} \frac{L_{\nu_b}}{R^3 \sigma_T} \frac{9}{2} \frac{3-p}{2-p} \left( \frac{3 e m_e c}{4 \pi} \right)^{1/2}.
\end{equation}

In Fig. \ref{fig:constraint}, we present the $B-R$ relation for a range of $\sigma$ values as solid lines in shades of green. We can see that high $\sigma$ values are not consistent with the self-absorption limit. A low $\sigma$ has also been inferred for Galactic PWNe, like Crab Nebula \citep{1984ApJ...283..710K}. Note that $\sigma=4/3$ corresponds to the minimum energy requirement, often used to infer the energetics of non-thermal radio sources \citep{2011hea..book.....L}.

Moreover, we can also see that the VLBI upper-limit to the source size limits the magnetic field to be below $0.05$~G.

The second y-axis of Fig. \ref{fig:constraint} correspond to the age of the source under the assumption that $\nu_b$ is due to synchrotron losses, given by,
\begin{equation}
t_{\rm age}=524.5 \left(\frac{\nu_b}{9.2{~\rm GHz}}\right)^{-1/2} \left(\frac{B}{0.01{~\rm G}}\right)^{-3/2} {~\rm yr}.    
\label{eq:ageeqn}
\end{equation} 
Cooling time corresponding to $B=0.05$~G is $42$ years, which gives a lower-limit to the age of the persistent source. 

We also see how limits on the average expansion velocity $<v>$ of the source can be represented on the $B-R$ plane. There are two constraints on the age of the source. The first one is from the break due to synchrotron cooling and the second one is from the radius ($t_{\rm age} = R/<v>$). Together, they lead to a $B$ vs $R$ relation for a given $<v>$. These are shown as dashed curves in shades of blue corresponding to different $<v>$.  As expected, higher velocity curves and higher $\sigma$ appear at the upper part of the plane. We see that, to be consistent with the self-absorption limit, $<v> < 0.025 c$.

In Fig. \ref{fig:constraint}, we also show the highest magnetic field possible, assuming that the entire rotational energy of a maximally spinning neutron star ($10^{52}$~ergs) is converted to the magnetic field of the plasma. 
\subsection{Numerical SED and MCMC parameter estimation}
\label{subsect32}
%----------- Two column figure (place early!)
   \begin{figure*}
   \centering
   \includegraphics[width=0.5\hsize]{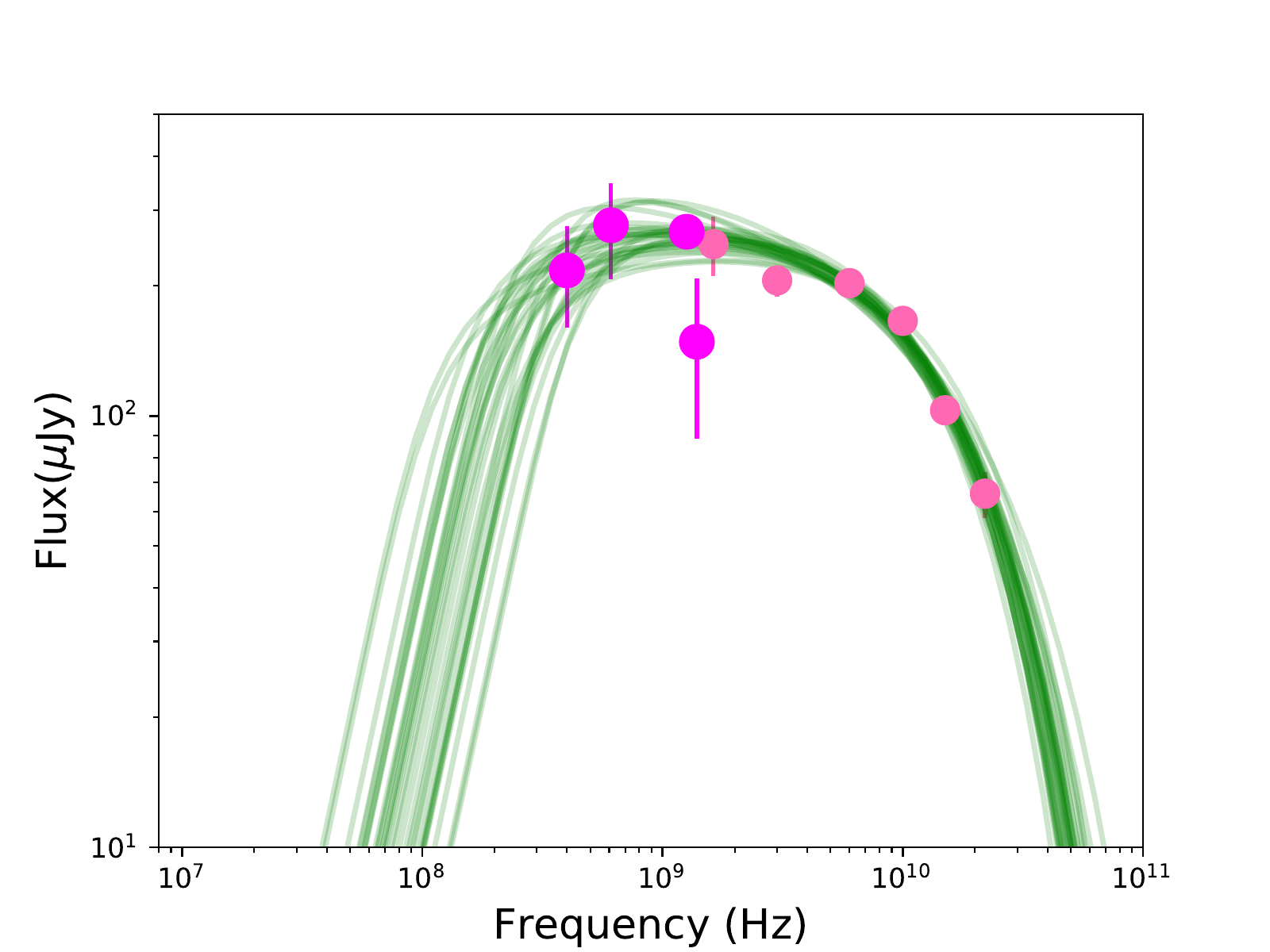}
   \includegraphics[width=0.35\hsize]{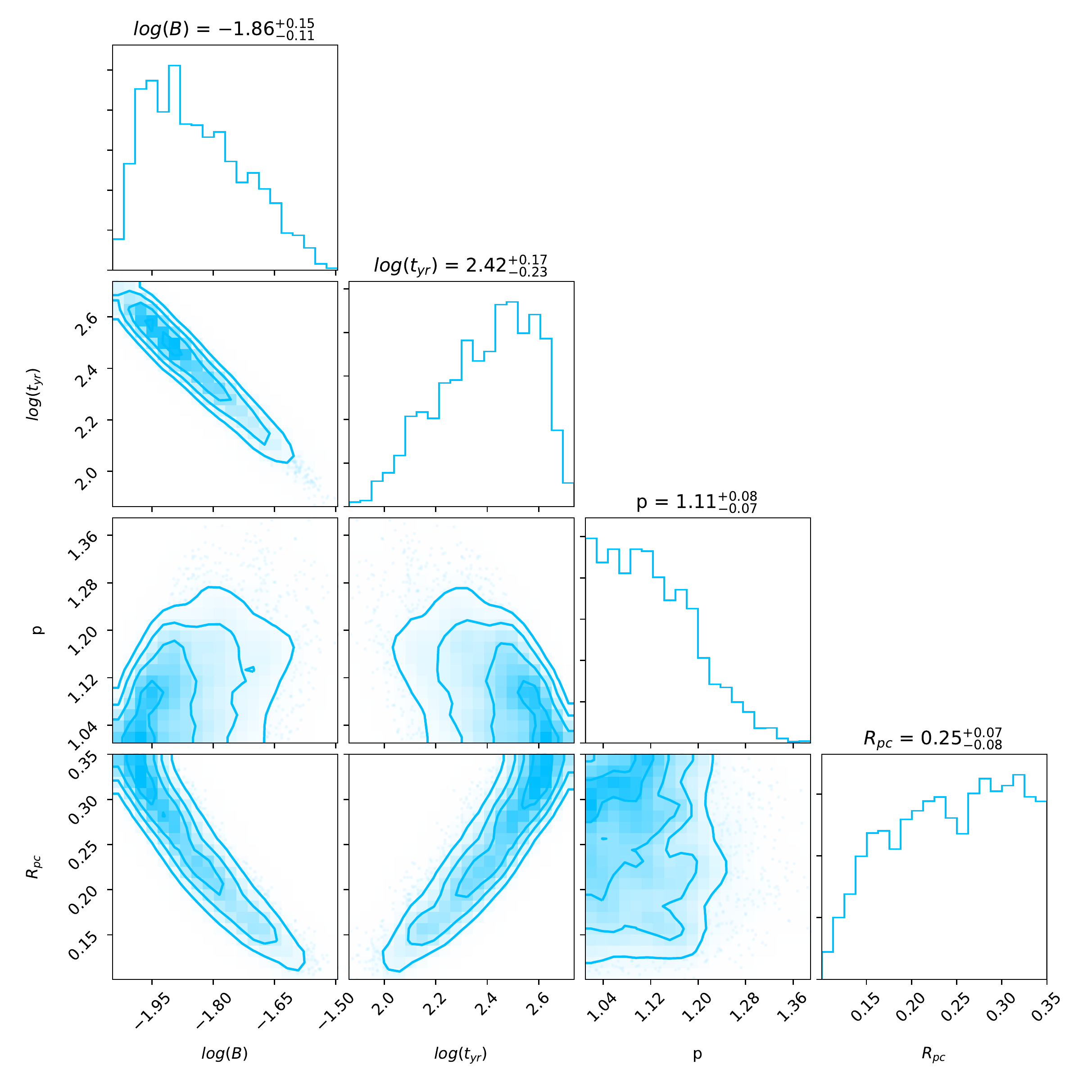}
      \caption{(left)Realizatoins of the SED from the Bayesian parameter estimation. (right) Posteriors of the parameter space given by $\log_{10}B, \log_{10}(t_{\rm yr}),p, R_{\rm pc}$.}
         \label{fig:numsed}
   \end{figure*}
To further understand the nature of the underlying electron distribution, we developed a synchrotron spectrum model by integrating the single electron power-spectrum over the electron distribution function $n(\gamma) d \gamma$. We considered a flat ($p <2$) electron distribution function affected by radiative cooling, given by
\begin{equation}
n(\gamma) =\frac{K_e m_e c^2}{(\gamma_m m_e c^2)^p} \left( \frac{\gamma}{\gamma_m} \right)^{-p} \left(1-\frac{\gamma}{\gamma_c} \right)^{p-2},
\label{elecool}
\end{equation}
where $K_e$ is the normalization in energy space and $\gamma_c$ is the break due to radiative losses \citep{JVbook}. In terms of the magnetic field $B$ in Gauss and the age $t_{\rm yr}$, $\gamma_c = \frac{24.5}{B^2 t_{\rm yr}}$ \citep{1979rpa..book.....R}.

We calculated the synchrotron emissivity $j_{\nu}$ and absorption coefficient $\alpha_{\nu}$ due to synchrotron self-absorption to finally arrive at the observed flux $f_{\nu}$ at a given frequency $\nu$ as $f_{\nu} = (\pi R^2/{d_L}^2) j_{\nu} \left( 1-\exp^{-\alpha_{\nu} R} \right)/\alpha_{\nu}$. The details of this calculation, following \citet{1979rpa..book.....R} is given in Appendix-B.

Ultimately, the parameter space determining the synchrotron spectrum is $\Theta = (R,B,t_{\rm yr},p,\gamma_m,\sigma)$. We fixed $\gamma_m = 10, \sigma = 4/3$ and performed a $4$ dimensional Bayesian parameter estimation using \textit{PyMultinest}, a nested-sampling algorithm \citep{2016ascl.soft06005B}. We find that the radius can not be very tightly constrained within the range of $<0.1 < R_{\rm pc} < 0.35$, while the other parameters have better limits. For the assumed $\sigma$, $B = 0.014^{+0.005}_{-0.004}$~G, $t_{\rm yr} = 263^{+163}_{-97}$, and $p=1.112 \pm 0.079$. In Fig. \ref{fig:numsed}, we present a few realizations from the posterior along with the data and the distribution of the parameter space.

%----------------------------------------------------------------- 

%-----------------------------------------------------------------
\subsection{Possible implications for the engine of the persistent source}
The analysis presented thus far puts constraints on magnetic-field strength and size of the progenitor, for a given magnetisation parameter.  These constraints provide also constraints on the total energy contained in the persistent source. For the total internal energy we have
\begin{equation}
E_{\rm int} \approx \left( u_{\rm e} + u_{\rm B}\right)\frac{4\pi}{3}R^3.
\end{equation}
Using the magnetisation parameter we can write
\begin{equation}
    u_{\rm e}+u_{\rm B}=\left(1+ \frac{1}{\sigma}\right) \frac{B^2}{8\pi}
\end{equation}
The analysis showed that $B\sim 0.01$~G, with $B>0.05$~G excluded, and values $B\ll 1$~mG requiring very low magnetisation values.
We, therefore, estimate the total energy in the source to be of order
\begin{equation}
E_{\rm int}\approx 2.1\times 10^{49} \left(1+ \frac{1}{\sigma}\right)\left(\frac{B}{0.01~{\rm G}}\right)^2\left(\frac{R}{0.35~{\rm pc}}\right)^3~{\rm erg}.
\end{equation}

The current paradigm for the origin of FRBs is that they are powered by NSs, potentially magnetars. For both normal pulsars and magnetars the source of energy for the persistent source is likely to come from the rotational energy of NS. The difference is, however, that the high surface magnetic-field  of a magnetar will result in a more rapid loss of the rotational energy.

The initial rotational energy of a NS is given by
\begin{equation}\label{eq:energy}
    E_{\rm rot,0}= \frac{1}{2}I \Omega_0^2 = 2.8\times 10^{50} \left(\frac{I}{1.4\times 10^{45}~{\rm g\ cm^2}}\right)\left(\frac{P_0}{10~{\rm ms}}\right)^{-2}~{\rm erg},
\end{equation}
with $I$ the NS momentum of inertia, $\Omega_0=2\pi/P_0$, the initial
rotation frequency, and $P_0$ the corresponding initial rotation period. 

From this expression we see that the initial rotation period needs to have been 
\begin{equation}
    P_0 \lesssim 36.5\left(1+ \frac{1}{\sigma}\right)^{-1/2}\left(\frac{B}{0.01~{\rm G}}\right)^{-1}\left(\frac{R}{0.35~{\rm pc}}\right)^{-3/2}~{\rm ms}.
    \label{eq:p0eqn}
\end{equation}
The nominal value for $P_0$ is not extremely short--- it is longer than the current period of the Crab pulsar. But note that
for $\sigma\ll 1$ or $R\ll 0.35$ smaller values for $P_0$ are required. Moreover, radiative energy losses and work done by the nebula on its surroundings may require a larger input energy than is
currently contained in the persistent source, and hence
would require a shorter initial period.

The time scale for the NS to lose this initial energy is typically
\begin{align} \label{eq:eqtau0}
    \tau_0 =& \frac{1}{2}\frac{P_0}{\dot{P}_0} \approx 8\times 10^2 \left( \frac{P_0}{10~{\rm ms}}\right)^2\left(\frac{B_{\rm p}}{10^{12}{~{\rm G}}}\right)^{-2}~{\rm yr}\\ \nonumber
    \approx& 10700 \left(1+ \frac{1}{\sigma}\right)^{-1}\left(\frac{B}{0.01~{\rm G}}\right)^{-2}\left(\frac{R}{0.35~{\rm pc}}\right)^{-3}\left(\frac{B_{\rm p}}{10^{12}{~{\rm G}}}\right)^{-2} \\ \nonumber 
    & \left(\frac{I}{1.4 \times 10^{45} {~\rm g cm^2}}\right) ~{\rm yr}.
\end{align}
 with $B_{\rm p}$ the magnetic-field strength at the poles of the neutron star. For this we substituted eq. \ref{eq:p0eqn} into the standard expression for $\tau_0$.
 
We already noted that the steepness of the radio spectrum beyond 10~GHz suggest not so much a continuous energy injection, but something that happened over a relatively short time scale compared to the age of the source. This suggest that in the context of a pulsar model $\tau_0 \ll t_{\rm age}$. This allows us to put constraints on the surface magnetic field of the NS. 
Using eqn. \ref{eq:eqtau0} for $\tau_0$ with eqn. \ref{eq:ageeqn} for $t_{\rm{age}}$, we get the following constraint:
\begin{equation}
\begin{split}
    B_{\rm p}&> 4.5\times 10^{12}\left(1+ \frac{1}{\sigma}\right)^{-1/2}\left(\frac{B}{0.01{~\rm G}}\right)^{-1/4} \left(\frac{R}{0.35{~\rm pc}}\right)^{-3/2} \\ 
   & \left(\frac{\nu_b}{9.2{~\rm GHz}}\right)^{1/4} \left(\frac{I}{1.4 \times 10^{45}{~\rm g cm^2}}\right)^{1/2} ~{\rm G}.
\end{split}
\end{equation}
This lower limit on $B_{\rm p}$ is consistent with values for normal young pulsars. So in principle a magnetar origin is not needed. But note that this conclusion strongly depends on the magnetisation parameter, and it should be noted that radiative energy losses, and work of the nebula on its surrounding may require a larger energy input than assumed here. It is nevertheless interesting that a magnetar as a central source is not necessarily required --- but also not ruled out. This may be surprising, but it should be noted that the total energy in the persistent source Eq.~\ref{eq:energy} is not that much different from the Crab Nebula, but the energy density is much higher due to the compactness of the source. 
Moreover, in our calculations we set $\tau_0$ equal
to the age of the source. But the steepness of the
spectral break may imply $\tau_0\ll t_{\rm age}$.
Setting, somewhat arbitrarily $\tau_0=0.1 t_{\rm age}$,
we would require a three times stronger pulsar magnetic field. 

\section{Conclusions}
\label{sec:end}
We present low frequency observations of the persistent counterpart of FRB121102 with the uGMRT in $400, 610, 1260,$ and $1390$~MHz frequencies. We detect an optically thin spectral energy distribution down to $400$MHz, with a flat spectral index similar to that of Galactic plerions. Using a generic synchrotron spectral model, we obtain constraints on the magnetic field density, radius, and age of the emitting plasma. We also constructed a numerical synchrotron SED and estimated the parameters through a Bayesian algorithm, and arrived at more robust constraints on $R, B, t_{\rm yr},$ and $p$.
Our conclusions are sensitive to the magnetisation parameter, $\sigma$, and also assume
that the spectral break at $\sim 10$~GHz is due to
radiative losses. With these assumptions in mind,
we list here our main conclusions:

\begin{itemize}
    \item Based on the absence of synchrotron self-absorption, and using energetic constraints, we arrive at upper limits for magnetic field and magnetization of the emitting region, $B<0.05$~G and $\sigma <100$ respectively. 
    \item For reasonable values of $\sigma$ the inferred magnetic field is $\sim 0.01$ G.
    \item Constraints on the age of the source are $t_{\rm age} > 524 \left( \frac{B}{0.01 {\rm G}}\right)^{-3/2}$, from assuming the spectral break at $9.2$~GHz to be due to radiative losses.
    \item We see that the emission region has a low $\sigma$ and is also expanding non-relativistically ($<0.025 c$). The slow expansion speed is consistent with the observed absence of systematic variations in the radio flux.
    \item Similar to Galactic PWNe, most of the energy in the persistent source is carried by leptons, and not by the magnetic field. This is similar to the Crab Nebula. However, this source has about three orders of magnitude stronger $B$-field in comparison with the Crab nebula. Moreover, it appears much younger, and the sharp break in frequency suggests a sharper decline in injection of electrons. 
    \item Assuming the rotational energy of the central neutron star to be responsible for the energy in the nebula, we obtained limits on its initial period to be shorter than $\sim 36$~ms.
    \item As the observed radio spectrum with its steep break implies a short energy injection time-scale, the characteristic age of the neutron star has to be smaller than the age of the nebula, and hence the limiting magnetic field of the neutron star $B_p > 4.5\times 10^{12}$~G. This result, obtained purely using the radio spectrum of the persistent source suggests that while a magnetar is not ruled out, it is also not necessarily required.
\end{itemize}
\begin{acknowledgements}
This work is partially funded by the Dept. of Science and Technology, India, grant EMR2016/007127. RL thanks J.W.T. Hessels, D. Bhattacharya, R. Dastidar, K. Misra, and Benjamin Stappers for stimulating discussions at various stages of this work. 
\end{acknowledgements}

\bibliographystyle{aa} % style aa.bst

 % your references Yourfile.bib

\begin{appendix}
\section{Normalization of the electron distribution}
In section-\ref{subsect31}, we have used the normalization $K_e$ of the electron distribution in terms of the observed luminosity at the break $\nu_b$. Below, we provide the derivation for the same. 

$K_e$ is defined through the density of electrons $n = K_e E^{-p}$, where $E$ is the energy of the electron given as $\gamma m_e c^2$.
 
The total luminosity of a synchrotron source, with an electron distribution extending from $\gamma_m$ to $\gamma_M$ can be written as 
\begin{equation}
L_{\rm syn} = V \int_{\gamma_m}^{\gamma_M} d \gamma P_{\rm syn}(\gamma) n(\gamma), 
\end{equation}
where $V$ is the volume of the source, $P_{\rm syn} (\gamma)$ is the synchrotron power radiated by an electron of Lorentz factor $\gamma$ and $n(\gamma)$ is the electron distribution function ($\int_{\gamma_m}^{\gamma_M} d \gamma n(\gamma)$ equals the number density of electrons). 

Using $V=(4/3) \pi R^3$, $P_{\rm syn} (\gamma) = \gamma^2 \frac{4}{3} c \sigma_T \frac{B^2}{8 \pi}$, one can rewrite the above equation as,
\begin{equation}
  L_{\rm syn} = \frac{2}{9} R^3 c \sigma_T  B^2 \frac{K_e \gamma_M^{(3-p)}}{(3-p) (m_e c^2)^{(p-1)}}. 
\end{equation}
For deriving this expression, we have once again assumed that $\gamma_M >> \gamma_m$ and ignored a $\gamma_m^{3-p}$ term in the integration.

The total luminosity $L_{\rm syn}$ can be approximately re-written as $L_{\nu_b} \nu_b$. With this substitution, one can arrive at equation-\ref{keeqn}.

The energy density in non-thermal electrons, $u_e \approx \frac{K_e}{2-p} (\gamma_M m_e c^2)^{(2-p)}$ can be re-written as,
\begin{equation}
u_e = L_{\nu_b} \nu_b^{1/2} \frac{9}{2} \frac{3-p}{2-p} \left( \frac{3 e m_e c}{4 \pi} \right)^{1/2} \frac{1}{B^{3/2} R^3 \sigma_T},
\end{equation}
after using $\nu_b = \nu_{\rm syn} (\gamma_M)$.

\section{Expressions used in section \ref{subsect32}}
In this section, we describe the steps followed in constructing the numerical synchrotron SED starting from the single electron power-spectrum. We have followed the method in \citet{1979rpa..book.....R}. The single electron power-spectrum $P_{\nu,\gamma}$ is $\frac{\sqrt{3} e^3 B}{m_e c^2}F(x)$ where $F(x)$ is $x \int_x^{\infty}  d\zeta K_{5/3} (\zeta)$, $K_{5/3}$ is the modified-Bessel function of $5/3$rd order, and $x$ is the normalized frequency $\nu/\nu_{\rm syn}(\gamma)$. We obtain the emissivity  $j_{\nu}$ by integrating the power per unit solid angle $P_{\nu, \gamma}/(4 \pi)$ with the electron distribution function given in \ref{elecool}. To estimate the optical depth $\tau_{\nu}$, we used the expression of the absorption coefficient $\alpha_{\nu} =\frac{\sqrt{3} e^3}{8 \pi m_e} \left( \frac{3 e}{2 \pi m_e^3 c^5} \right)^{p/2} K_e \Gamma\left[ \frac{ep+22}{12}\right] \Gamma\left[ \frac{3p+2}{12}\right] B^{(p+2)/2} \nu^{-(p+4)/2}$. Normalization $K_e$ of the electron energy distribution function is re-written in terms of the energy density $u_e$ in non-thermal electrons as $K_e = (2-p) u_e \left(\gamma_M m_e c^2 \right)^{p-2}$. As we have assumed in section-\ref{subsect31}, we then considered the energy densities $u_e$ and $u_B$ to be related through an arbitrary magnetization parameter $\sigma$ to finally write $\alpha_{\nu}$ in terms of $B, \gamma_M, p,$ and $\sigma$. The final observed flux is calculated as $f_{\nu} = \frac{\pi R^2}{{d_L}^2} \frac{j_{\nu} (1-\exp{(-\alpha_{\nu}R)})}{\alpha_{\nu}}$.
\end{appendix}

\end{document}